\documentclass[twocolumn, prl]{revtex4}
\usepackage{graphicx}
\usepackage{epsfig}
\begin{document}
\title{$^{52}$Cr Spinor Condensate$ -- $ A Biaxial or Uniaxial Spin Nematic}
\author{Roberto B. Diener and Tin-Lun Ho}
\affiliation{Department of Physics, The Ohio State University, Columbus, OH}
\date{\today}
\begin{abstract}
%We show that the newly discovered Bose condensate of $^{52}$Cr is either a uniaxial or a biaxial spin nematics in the limit of weak magnetic fields, depending on an unknown scattering length of the system. Our conclusions are draw from the exact solution of the general phase 
We show that the newly discovered $^{52}$Cr Bose condensate in zero magnetic field can be a spin nematic of the following kind: A ``maximum" polar state, a ``co-linear" polar state, or a biaxial nematic ferromagnetic state. We also present the phase diagram with a magnetic field in the interaction subspace containing the Chromium condensate.  It contains many uniaxial and biaxial spin nematic phases, which often but not always break time reversal symmetry, and can exist with or without spontaneous magnetization.
%
%We show that depending on the scattering length in the singlet channel, the newly discovered $^{52}$Cr Bose condensate can be one of the following three spin nematic state: A ``maximum" polar state, a ``co-linear" polar state, and a biaxial nematic ferromagnetic state.  Our result is based on the solution of the global phase diagram of a spin-3 Bose condensate.  It contains many uniaxial or biaxial spin nematic phases, which often but not always break time reversal symmetry, and can exist with or without spontaneous magnetization.% There are strong  connections between magnetization (spontaneous or induced) and biaxial nematicity.  
\end{abstract}
\maketitle

Recently,  Tilman Pfau's group at Stuttgart has succeeded in condensing a Bose gas of $^{52}$Cr\cite{Pfau}. $^{52}$Cr atoms are spin-3 bosons with electronic spin $J=3$ and nuclear spin $I=0$.  In contrast, alkali bosons such as $^{87}$Rb and $^{23}$Na are spin-1 or -2 bosons with $J=1/2$ and $I=3/2$.  Since  $^{52}$Cr has a $J$ 6 times larger than the alkali's, its dipolar interaction is 36 times larger.  Indeed, dipolar effects have been observed in Cr\cite{scattering},  although they have not been observed in the alkalis.  For this reason, Cr is sometimes referred to as  a ``dipolar condensate. This characterization is not quite appropriate. Despite their observability, the dipolar energy is no more than a few present of the total energy.  As we shall see,  if the spin degrees of freedom of Cr is released, it will have many remarkable spin nematic structures stabilized by energies much larger than the dipole energy. The Cr condensate, in short, is a quantum spin nematic. 

At present, experiments on $^{52}$Cr are performed in magnetic traps which freeze the atomic spins.  The spin degrees of freedom, however, can be released in optical traps.  
%As a result, the atoms behave like spinless rather than spin-3 bosons.  As in the case of alkali condensates, the true spin nature of the Cr condensate can be revealed in optical traps, which trap all spin states.   
In the case of optically trapped  $^{87}$Rb and $^{23}$Na, %in optical traps, 
ground states of different magnetic structures have been discovered. 
Since Cr atoms have a higher spin,  the number of possible phases will increase.  Cr also differs from the alkalis in that  its scattering lengths in different angular momentum channels are very different, whereas they are very similar in the alkalis. According to ref.\cite{Ho}, the closer these scattering lengths, the weaker the spin-dependent interactions. 
Thus the ratio between spin-dependent and density-density interactions in the alkalis is about $1\%$ while it is of order 1 in  $^{52}$Cr.  In other words, once the spins of 
Cr are unfrozen,  {\em the spin-dependent interactions will overwhelm the dipole energy}. Any novel properties of Cr spinor BEC will be consequences of the spin interaction with dipolar effects as a perturbation.

For spin-3 bosons, the interactions are specified by the scattering lengths $a_{S}$ in total angular momentum channel $S=0,2,4, 6$.   At present, all $a_{S}$ except $a_0$ have been determined\cite{scattering}.  The physical system therefore lies on a line in the interaction parameter space, which we refer to as the ``Cr"-line.  While we shall focus on this line, we also present the phase diagram as a function of interaction and magnetic field, for it will be useful for other spin-3 Bose systems. 
As we shall see,  the phase diagram is full of spin nematic phases which are either unaxial or biaxial.  These phases typically but not always break time reversal symmetry, which can come about 
with and without a spontaneous magnetization \cite{SantosPfau}.  The discovery of some of these nematic phases will be exciting indeed.

%Our result, however, are in complete disagreement with the recent preprint of  Santos and Pfau\cite{Santos}. 

{\bf I. The spin-dependent energy functional:}  Let $\Psi_{m}$$=$ $\sqrt{n}$ $\zeta_{m}$, $m= -3$ to  $3$, be the condensate of a spin-3 Bose gas, where $n$ is the density, and $\zeta^{\dagger} \zeta = 1$.  
% $\langle \psi_{\mu} \rangle$ 
%Let $\zeta$ be a normalized spinor of a spin-3 Bose condensate.  Its spin components will be denoted as  $\zeta_{m}$, $m=-3$ to $3$. 
As shown in ref.\cite{Ho}, the ground state of a homogenous or single mode spinor condensate in zero magnetic field is determined by the energy  
%\begin{equation}
$E/E_{o}$ $ = $ $\sum_{S=0,2,4,6}$ $ \tilde{a}_{S}{\cal P}_{S}^{}$,
% \,\,\,\, \,\,\,\,\,\,  
$\tilde{a}_{S}= a_{S}/a_{B}$, 
%\label{E} \end{equation} 
where $E_{o}$ $ = $ $\int  $ $(2\pi \hbar^2 a_{B}/M)n^2$, $a_{B}$ is the Bohr radius, 
${\cal P}_{S}$ $ =$ $  \sum_{M=-S}^{S}  |B_{S,M}|^2$, and 
%and $B_{S,M}$ is the amplitude of combining a pair spin-3 bosons into a total spin state $|S,M\rangl$, 
%\begin{equation}
%${\cal P}_{S} = |B_{S,M}|^2$, 
%\end{equation}
%where 
$B_{S,M}$ $ = $ $\sum_{m, m'} $  $ \zeta_{m} \zeta_{m'}$ $ \langle  S,M| 3, m; 3, m'\rangle$ is the amplitude of forming a boson pair in state state $|S, M\rangle$.  In the case of spin-1 alkali gases, (where $S=0,2$ in the energy $E$), the $a_{S}$'s differ from each other only by about a few percent, so $E$ is almost an identity. In contrast, $^{52}$Cr has $a_2 = -7a_{B}$, $a_{4} = 58 a_{B}$, $a_{6} = 112a_{B}$, while $a_{0}$ is undermined\cite{scattering}.  
%$E$ is far from being an identity. 
Using the method in ref. \cite{Ho}, 
%\end{equation}
% eq.(\ref{E}) 
 $E$ can be written as\cite{cal1}  
\begin{equation}
E=E_{o}\left[  \alpha |\Theta|^2 + \beta \sum_{M=-2}^2 |B_{M}|^2 + \gamma \langle {\bf S}\rangle \cdot  \langle {\bf S}\rangle  + C\right] 
\label{energy}\end{equation}
%Using the identities $1 = P_{0} + P_{2} + P_{4} + P_{6}$, ${\bf S}_{1}\cdot{\bf S}_{2} = -12 P_{0} -9 P_{2} -2 P_{4} + 9P_{6}$$=9 -21P_{0} - 18P_{2} - 11P_{4}$ for a pair of spin-3 bosons with contact interaction,  where $P_{S}$ is the projection operator for total spin $S$,  we can trade $P_{6}$ and $P_{4}$ with  a constant and ${\bf S}_{1}\cdot {\bf S}_{2}$. In this way, eq.(\ref{E}) can be written as 
%\begin{equation} E=E_{o}\left[  \alpha |\Theta|^2 + \beta \sum_{M=-2}^2 |B_{M}|^2 + \gamma \langle {\bf S}\rangle \cdot  \langle {\bf S}\rangle  + C\right]   \label{energy}\end{equation}
where 
$\langle {\bf S}\rangle = \zeta^{\ast}_{m} {\bf S}_{m m'}^{}\zeta^{}_{m'}$,   $B_{M} = \sqrt{7} B_{2,M}$, $\Theta = \sqrt{7} B_{0,0}$, so that ${\cal P}_{2}^{} = 7
\sum_{M}|B_{M}|^2$ and ${\cal P}_{0}^{} = 7 |\Theta|^2$. 
The coefficients are defined as 
$\alpha$ $=$ ${1\over 7 a_{B}} $ $ [ (a_0-a_6) + {21\over 11} (a_6-a_4) ]$, 
$\beta$ $= $ $ {1\over 7a_{B}}$ $[(a_2-a_6) + {18\over 11} (a_6-a_4)  ]$, 
$\gamma $ $= $ $ {1 \over 11 a_{B}} (a_6-a_4)$, and $C= {1\over 7 a_{B}} $ 
$[ a_6 - {9\over 11}(a_{6} - a_{4})]$.  We then have $\beta = -4.38$, $\gamma=4.91$, $C= 67.8$, and $\alpha$ undetermined. 
 Using the relation between ${\bf S}_{1}\cdot{\bf S}_{2}$ , $({\bf S}_{1}\cdot{\bf S}_{2})^2$ and the projection operators $P_{i}$\cite{SS2}, 
we can further express eq.(\ref{energy}) in terms of the  nematic tensor ${\cal N}_{ij}^{} \equiv \langle (S_{i}^{} S_{j}^{} + S_{j}^{} S_{i}^{})\rangle/2$ as 
\begin{equation}
E/E_{o}= \overline{\alpha} |\Theta|^2 + \overline{ \beta} {\rm Tr}{\cal N}^2 + 
\overline{\gamma} \langle {\bf S}\rangle \cdot  \langle {\bf S}\rangle + \overline{C},
\label{energynew}\end{equation}
where $\overline{\alpha} = \alpha - {5 \over 3}\beta$, $\overline{\beta} = \beta/18$,  $\overline{\gamma}
=\gamma -  {5\over 12}\beta $, $\overline{C} = C - 10\beta$.  For Cr, we have  $\overline{\beta} = - 4.38$, $\overline{\gamma} = 2.36$, $\overline{C} = 111.56$ and $\overline{\alpha}$ is unknown. 

{\bf II. Dipole energy :}   
To get an idea of the general structure of $E_{D}$, we use the single mode approximation so that 
all components of $\Psi_{\mu}$ have the same spatial function, 
$\Psi_{\mu}({\bf r}) = \sqrt{n({\bf r})} \zeta_{\mu}$, and $\zeta_{\mu}$ has no spatial dependence. For harmonic traps $V_{T} = \frac{M}{2}\sum_{i=1}^{3}\omega_{i}^2( {\bf r}\cdot\hat{\bf a}_{i})^2$, 
where $(\hat{\bf a}_{1}, \hat{\bf a}_{2}, \hat{\bf a}_{3})$ is %a set of orthogonal 
an orthonormal triad, the dipole energy is (see footnote~\cite{dipole})
\begin{equation}
E_{D}/E_{o}= \sum_{i=1}^{3} \eta_{i}^{} (\langle {\bf S}\rangle\cdot \hat{\bf a}_{i})^2, 
\label{EDnew} \end{equation}
where $\eta_{i} =  \frac{1}{2E_{o}}(g\mu_{B})^2\int {\rm d}{\bf r}_{1}{\rm d}{\bf r}_{2} n({\bf r}_{1}) n({\bf r}_{2})
[1 - 3(\hat{\bf a}_{i} \cdot \hat{\bf r})^2]/r^3$, ${\bf r} = {\bf r}_{1}- {\bf r}_{2}$.  In particular, we denote the direction with smallest $\eta$ as $\hat{\bf a}_{3}$. Combining the $\gamma$ term in eq.(\ref{energy}) with eq.(\ref{EDnew}), all $\langle {\bf S}\rangle$-dependent terms in the energy can be written as in eq.(\ref{EDnew})  with $\eta_{i}\rightarrow  \eta_{i} + \gamma$. The effect of the dipole energy is to align $\langle {\bf S}\rangle$ along the direction $\hat{\bf a}_{3}$. 
The phase diagram in zero magnetic field including dipole energy can therefore be obtained by first minimizing eq.(\ref{energy}) (without dipole energy), then replace $\gamma$ by $\gamma - \eta_{3}$, and align the spontaneous magnetization of those phases that acquire it along $\hat{\bf a}_{3}^{}$.

To estimate the magnitude of $\eta_{i}$, it is straightforward to show that 
%\begin{equation}
$\eta_{i} =  \frac{ -( g\mu_{B})^2}{ 4\pi\hbar^2 a_{B}/M}\frac{4\pi I_{i}}{3} 
%=  \frac{- g^2}{4} \left(\frac{e^2}{\hbar c}\right)^2 \frac{M}{m_{e}} \frac{I_{i}}{3} 
= -1.7 I_{i}  $
%\label{eta} \end{equation}
where $ I_{i} = \int_{\bf q} |n({\bf q})|^2 (1 - 3( \hat{\bf a}_{i}\cdot \hat{\bf q})^2)/ \int_{\bf q} |n({\bf q})|^2$.
%, and we have used $g=2$, $e/\hbar c = 1/137$, $M/m_{e}= 52 \times 1836$.  
Approximating 
 $n({\bf r})$  as a Gaussian so that $n({\bf q}) = C e^{- \sum_{i=1}^{3}q_{i}^2 R_{i}^2}$, where $R_{i}$ is the radius of the atom cloud along
$\hat{\bf a}_{i}$and
% . It is then easy to show that f
for cylindrical traps ($R_{x}=R_{y} = R_{\perp} \neq R_{z}$), 
$I_{z} $ varies from 0 to 0.7 as  $R_{z}/R_{\perp}$ varies from 1 to 10. Thus, $\eta_{i}$ is at most about 25$\%$ of $\gamma$. 
As we shall see, all the phases along the Cr-line either have zero magnetization, or a weak spontaneous magnetization such that the $\gamma$ term in eq.(\ref{energy}) contributes very little to the total energy.  Since $E_D$ is dominated by the  $\gamma$ term,  it can be ignored as a first approximation.

{\bf III. Relation between singlet amplitude $\Theta$, magnetization $\langle {\bf S}\rangle$, and the nematic tensor ${\cal N}$:}   The quantities $\Theta$, $\langle {\bf S}\rangle$, and ${\cal N}$ provide a characterization of the phases.   They are, however, not independent quantities. 
%Before proceeding, we first point out some important properties of  $\Theta$, $\langle {\bf S}\rangle$, and ${\cal N}$, and their relations with each other. These properties and relations provide us a characterization of the phases.  We 
Let us first consider $\Theta$, which is 
%\begin{equation}
$\Theta = 2\zeta_3 \zeta_{-3} - 2\zeta_2 \zeta_{-2}+2\zeta_1 \zeta_{-1} -\zeta_0^2 \equiv A^{(o)}_{m m'}\zeta_{m}^{}\zeta^{}_{m'}$, 
%\end{equation} 
where  $A^{(o)}_{m m'}$ $ = $ $(-1)^{m+1}$ $ \delta_{m+m',0}$ is the matrix for spin change 
under time reversal.  It is also easy to show that  the maximum value of $|\Theta|^2$ is 1, and the condition for this is that $\zeta$ is invariant (up to a phase $\chi$) under time reversal, i.e. 
%\begin{equation}
% \zeta_{m}^{\ast} = (-1)^{m+1}e^{i\chi} \zeta_{-m}^{}, 
$(A^{(o)}\zeta)^{\ast}_{m} = e^{i\chi}\zeta_{m}^{}$.
%if 
% \,\,\,\,\,\,\,\, {\rm if} \,\,\,\,\,\,\,\, 
%$ |\Theta |=1.  $
%\label{trs} \end{equation}
This also implies $\langle {\bf S} \rangle=0$~\cite{spin-2}.%which follows from the identity   $A^{(o)}{\bf S} + {\bf S}^{T}A^{(o)}=0$,  where the superscript $``T"$ means transpose.  
%The latter condition is a consequence of the rotational invariance of $\Theta$. 
Thus {\em any state with $|\Theta|<1$ breaks time reversal symmetry. }

 In the presence of a magnetic field ${\bf B}= B\hat{\bf z}$, the rotational invariance of 
 eq.(\ref{energy})  implies that 
 % it is useful to establish a set of Cartesian axis ($\hat{\bf x}, \hat{\bf y}, \hat{\bf z}$) such that $\hat{\bf B}= \hat{\bf z}$. Since the energy eq.(\ref{???}) is rotational invariance, a magnetic field ${\bf B}$  will generate a magnetization $\langle {\bf S}\rangle$ along along $\hat{\bf z}$. The energy in the presence of field is given by 
\begin{equation}
E/E_{o}= \alpha |\Theta|^2 + \beta \sum_{M=-2}^2 |B_{M}|^2 + \gamma \langle {\bf S}^{} \rangle^2   -  p\langle S_{z}^{}\rangle + C,
\label{EB}\end{equation}
where $p = g\mu_{B}B/E_{o}$. 
Next, we note that the tensor ${\cal N}$ is hermitian. It then has the diagonal form
%\begin{equation}
${\cal N}_{ij} = \lambda_{1}\hat{\bf e}_{1i} \hat{\bf e}_{1j} + \lambda_{2}\hat{\bf e}_{2i} \hat{\bf e}_{2j} + \lambda_{3}\hat{\bf e}_{3i} \hat{\bf e}_{3j} $, 
%\end{equation} 
where $\{ \hat{\bf e}_{a} \}$ are the principal axes, ($a=1,2,3$), $\hat{\bf e}_{a} \cdot \hat{\bf e}_{b} = \delta_{ab}$, and the $\lambda_{a}= \langle (\hat{\bf e}_{a}\cdot {\bf S})^2\rangle>0$  are the eigenvalues, satsifying
\begin{equation}
\lambda_{1} + \lambda_{2} + \lambda_{3} = S(S+1) = 12,\,\,\,\,\,\,\,\,\, S=3.
\label{lambdasum} \end{equation}
We shall order the $\lambda$'s so that $\lambda_{1}\leq \lambda_{2} \leq \lambda_{3}$. The axes
 $(\hat{\bf e}_{1}$, $\hat{\bf e}_{2}$, $\hat{\bf e}_{3})$ will be referred to as minor, middle, and major axis respectively.  Without loss of generality, we can always arrange  $(\hat{\bf e}_{1}, \hat{\bf e}_{2} , \hat{\bf e}_{3})$ in a right handed triad.  Using the terminology of liquid crystals, the system is referred as a (spin) nematic if ${\cal N}$ is not isotropic, i.e. not proportional to the identity matrix. Systems with two identical eigenvalues and three 
unequal eigenvalues will be referred to as  uniaxial and biaxial nematics respectively.   

%Note that the quantities  $\Theta$, $\langle{\bf S}\rangle$, and ${\cal N}$ are not independent of each other, and their competition in eq.(\ref{energynew}) is the reason for different phases. 

The origin of different phases is due the competition of $\Theta$, $\langle{\bf S}\rangle$, and ${\cal N}$, 
which are not independent quantities. Such competition can be illustrated by considering $\overline{\beta}<0$,  in which case ${\rm Tr} {\cal N}^2$ in eq.(\ref{energynew})  favors $\lambda_{3} = 9$, $\lambda_{1}=\lambda_{2} = 3/2$, which can be achieved by either the polar state $(1,0,0,0,0,0,1)/\sqrt{2}$, or the ferromagnetic state $(1,0,0,0,0,0,0)$.  
However, neither of these states are favored simultaneously by the $\Theta$ term when $\overline{\alpha}>0$, and the $\langle {\bf S}\rangle^2$ term, since $\overline{\gamma}>0$  for Cr.  One of the key features we find below is that magnetization (be it spontaneous or induced) is generally accompanied with biaxial nematicity.  Even though the energy functional is rotationally symmetric about the external field $\hat{\bf z}$, this symmetry is broken spontaneously in the biaxial nematic state.

\begin{figure}[t]
\centering \epsfig{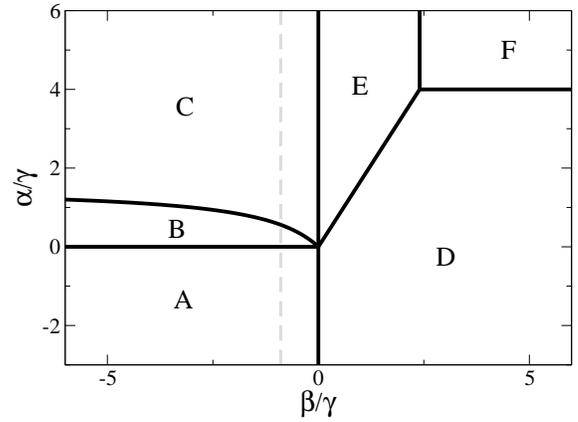}
\caption{Phase diagram in zero magnetic field for $\gamma>0$.  The dashed line (referred to as the ``Cr"-line) is where the physical system lies. The solid lines are first order boundaries. The nature of the phases displayed are discussed in Fig. 3.} \label{phase_diagram.fig}
\end{figure}

{\bf IV. Phase diagram in zero magnetic field, $p=0$:} 
We have minimized eq.(\ref{energy}) by a combination of analytic and numerical methods.  The phase diagrams are shown in fig.1 and 2. 
%All phase boundaries are straight lines except those separating $B$-$C$ and  $G$-$H$.   Boundaries %of first and second order phase transition are represented by black and grey lines respectively. 
Since $\alpha$ is unknown for $^{52}$Cr, the physical system lies on the dotted straight line (or ``Cr"-line for short)  in fig.1, $\beta/\gamma = -0.892$. It passes through the phases $A$, $B$, and $C$, which we name ``maximum polar" state (A), co-linear polar state (B), and biaxial nematic ferromagnetic state (C), respectively.   All order parameters $\zeta$ displayed below are unique up to a phase factor and an arbitrary spin rotation. 
% If an infinitesimal magnetic field  (along $\hat{\bf z}$) is present, rotational degeneracy is 
%reduced  to arbitrary rotation about $\hat{\bf z}$. 

% For both $\gamma>0$ and $\gamma<0$,  The nematic tensor ${\cal N}$ of the ground state has an axis aligned with $\hat{\bf B}$,  so that we have $(\hat{\bf e}^{(1)}, \hat{\bf e}^{(2)}, \hat{\bf e}^{(3)}) = (\hat{\bf x}, \hat{\bf y}, \hat{\bf z})$.  However, this is not always for non-zero magnetic fields, $p\neq 0$. 
%In the following, we shall discuss the states $A, B$ and $C$ in detail. The notations we set up will be used to describe all other phases.  

{\em Symbols and notations:}  We characterize each phase by its condensate wavefunction $\zeta$, singlet amplitude $\Theta$, magnetization $\langle {\bf S} \rangle$ and nematic tensor ${\cal N}$. An isotropic ${\cal N}$ ($\lambda_{1}=\lambda_{2}=\lambda_{3}$) will be denoted by a sphere.  Uniaxial nematics with $(\lambda_{1}= \lambda_{2}< \lambda_{3})$ and $(\lambda_{1}< \lambda_{2}= \lambda_{3})$ are represented by a long and flat cylinder, with major and minor axis being the symmetry axis of the cylinder respectively.  Biaxial nematics with $(\lambda_{1}< \lambda_{2}< \lambda_{3})$ are represented as a ``brick" with edge lengths given by $\lambda_{a}$. The principal axis $\hat{\bf e}_{a}$ is parallel to the edge with length $\lambda_{a}$. Figure 3 shows the structure of the phases.  In particular,
%Since the phases $(A), (B)$ and $(C)$ are the only candidates of the Cr condensate, we discuss them particularly.  The eigenvalues $\lambda_{a}$ and the magnetization $\langle {\bf S}\rangle$ of these phases along the Cr-line are shown in figure 3 and 4 respectively.  The properties of these phases are 
\begin{figure}[t]
\centering \epsfig{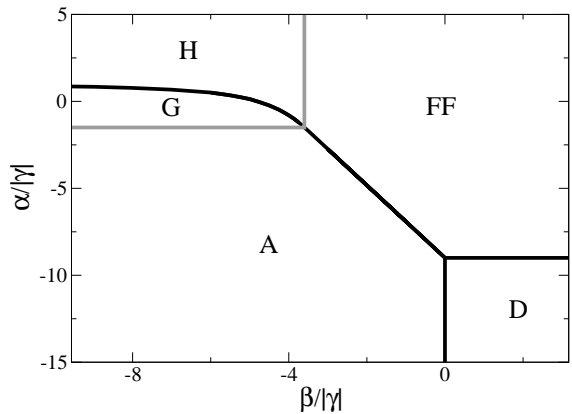}
\caption{Phase diagram in zero magnetic field for $\gamma<0$.  The solid and grey lines are first and second order boundaries, respectively. The nature of the phases is discussed in Fig. 3.} \label{phase_diagram_neg.fig}
\end{figure}
\begin{figure}[t]
\centering \epsfig{file=Slide1.epsf, width=3.2in}
\caption{Phases present at $p=0$.  $a,b,c,d$ and $\delta$ are real numbers. } \label{Table1.fig}
\end{figure}
\begin{figure}[b]
\centering \epsfig{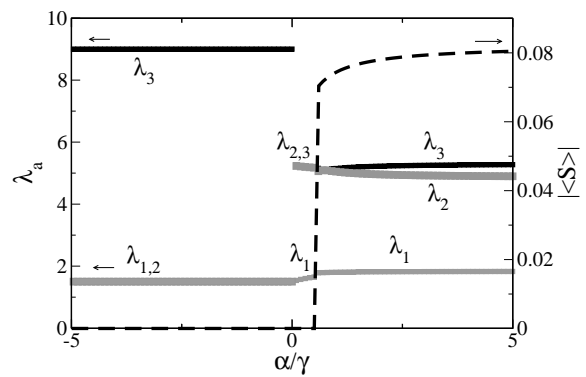}
\caption{The eigenvalues $\lambda_{a}$ of ${\cal N}$ and the magnetization $\langle {\bf S} \rangle$ along the Cr-line in figure 1.} \label{lambdas.fig}
\end{figure}
\noindent ${\bf (A)}$ {\em Maximum polar state}:  This state has time reversal symmetry, since $\Theta= 1$. It  is a uniaxial nematic with $\lambda_{3} = 9$, $\lambda_{1,2} = 3/2$. 
%\noindent 
${\bf (B)}$ {\em Co-linear polar state}:  It is a superposition of two polar states $(1,0,0,0,0,0,1)$ and $(0,0,1,0,1,0, 0)$ with unequal weight and a relative phase.  This system is  non-magnetic $\langle {\bf S}\rangle =0$ and yet has broken time reversal symmtry,  $0< |\Theta|< 1$. 
The latter is achieved by the phase angle $\delta$, which varies from 1.3 to 1.4 along the Cr-line from bottom to top. The system is uniaxial nematic with $\lambda_1< \lambda_2 = \lambda_{3}$. 
%\noindent 
${\bf (C)}$  {\em Biaxial nematic ferromagnet} : The system has a weak {\em spontaneous magnetization} $\langle {\bf S}\rangle = M\hat{\bf z}$,  $M=0.07$ to 0.08,  and is a biaxial nematic with either $\hat{\bf e}_{2}$ or $\hat{\bf e}_{3}$ along $\langle {\bf S}\rangle$.  Since $M$ is small, the contribution of the $\gamma$ term in eq.(\ref{energy}), (and hence dipolar energy) is a very small contribution to the total energy. This structure is driven largely by the competition between the nematic and the singlet contribution. The boundary separating $B$ and $C$ is  $\tilde{\alpha}   = 3|\tilde{\beta}|/ (3|\tilde{\beta}|+ 2), \,\,\,\,\,\,\,\, \tilde{\alpha} = \alpha/\gamma,  \,\,\,\,\,\,\,\, \tilde{\beta} = \beta/\gamma$.   The behaviors of $\lambda_{a}$ and $|\langle {\bf S}\rangle|$ are shown in fig.4. 
%\begin{equation} \overline{\alpha}   = \frac{3|\overline{\beta}}{3|\overline{\beta}|+ 2}, \,\,\,\,\,\,\,\, \overline{\alpha} = \frac{\alpha}{\gamma},  \,\,\,\,\,\,\,\, \overline{\beta} = \frac{\beta}{\gamma}. \end{equation}
%\begin{equation} \frac{\alpha}{\gamma}   = \frac{3\left| \frac{\beta}{ \gamma} \right|}{3\left| \frac{\beta}{ \gamma} \right|+ 2}. \end{equation}
The properties of these phases %in fig.1 and 2 
are tabulated in fig.4.  
%The numbers $a, b, c, d$ are real.   
The boundary between $G$ and $H$ is $\tilde{\alpha}  =( 24 \tilde{\beta} + 5 \tilde{\beta}^2)/
( 36 + 24 \tilde{\beta}  + 5 \tilde{\beta} ^2)$~\cite{I phase}. % For the $G$-phase, $\hat{\bf e}_{1} \parallel \langle {\bf S} \rangle$. %$(FF)$ means fully ferromagnetic. 
%While all order parameters are degenerate an infinitesimal magnetic field along $\hat{\bf z}$. 

{\bf V. Phase diagram in non-zero magnetic field:} 
%The phase diagram of S=3 spinor condensate in non-zero magnetic field has a surprisingly  rich structure. 
Figure 5 shows the phase diagram along the Cr-line in figure 1 for $p\neq 0$ obtained by minimizing eq.(\ref{EB}). It has an intricate  structure near $\alpha=0$.  All states have non-zero magnetization $\langle {\bf S}\rangle$ along the direction of the magnetic field  $\hat{\bf z}$, represented as an arrow. 
 %The numbers $a, b, c, d$ are real.   
The main features of our results are: 

\begin{figure}[t]
\centering \epsfig{file=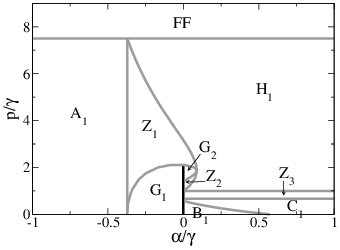, width=3in}
\centering \epsfig{file=Slide2.epsf, width=3.2in}
\caption{Top panel: Phase diagram for non-zero magnetic field along the Cr-line in fig.1.  The solid and grey lines are first and second order boundaries, respectively.  Bottom panel: Explanation of the phases in the diagram.  The numbers $a, b,c, d$ are real. All entries of the wavefunctions for the $Z_j$ phases are generally non-zero. It is not clear they can be represented in a simple form as the other phases.} \label{phase_diagram_p.fig}
\end{figure}

%
%\begin{figure}[t]
%\centering \epsfig{file=phase_diagram_p.eps, width=3in}
%\caption{Phase diagram for non-zero magnetic field along the Cr-line in fig.1.  The solid and grey lines are first and second order boundaries, respectively.} \label{phase_diagram_p.fig}
%\end{figure}

%\begin{figure}[t]
%\centering \epsfig{file=Slide2.epsf, width=3.2in}
%\caption{Phases present in a magnetic field.} \label{Table2.fig}
%\end{figure}

\noindent {\bf (i)} The transition $A_{1}\rightarrow Z_{1} \rightarrow G_{1}$ is 
%basically 
a rotation of uniaxial nematic tensor in the $A_{1}$ phase (represented as a long cylinder)  along its middle principal axis $\hat{\bf e}_{2}$  as one crosses the phase $Z_{1}$, with the rotation angle
finally reaching $90^{o}$ at the $G_{1}$ phase.  However, in the $Z_{1}$ phase, ${\cal N}$ acquires biaxial nematicity, which continues into the $G_{1}$ phase. 
%\noindent
 {\bf (ii)}   As $p\rightarrow 0$, for  $\alpha/\gamma< -0.372$
 and  $\alpha/\gamma> -0.372$, the states  $\zeta^{T}_{A_{1}}$ and $\zeta^{T}_{G_{1}}$ reduce to 
 $ (1,0,0,0,0,0,1)/\sqrt{2}$ and $(0,\sqrt{3},0,\sqrt{10},0,\sqrt{3},0)/4$ resp.,  which are related to each other by a $90^{o}$ rotation about $\hat{\bf y}$.  
%\noindent
 {\bf (iii)}  The transition  $A_{1}\rightarrow Z_{1} \rightarrow H_{1}$  is a similar rotation process as in ${\bf (i)}$ except that the % net  
 rotational angle  is zero when one reaches the $H_{1}$ phase. The nematic tensor ${\cal N}$ of $H_{1}$ is again uniaxial, except that $H_{1}$ has zero singlet amplitude, unlike $A_{1}$. 
%\noindent
 {\bf (iv)} $G_{1}$ and $G_{2}$ have similar tensor ${\cal N}$. There is, however, a jump in $\Theta$ across their boundary. 
% \noindent
{\bf (v)} As the $B$ phase in figure 1 extends into the $B_{1}$ phase in finite field $p$, its nematic tensor ${\cal N}$ changes from uniaxial to biaxial, merging into the $C_{1}$ phase, the finite field extension of $C$. %In both $B_{1}$ and $C_{1}$, the major axis $\hat{\bf e}_{1}$ of ${\cal N}$ is aligned with $\hat{\bf z}$. The difference between $B_{1}$ and $C_{1}$ is  ???? . 
%\noindent 
{\bf (vi)} Among all phases in finite field, $Z_{1}, Z_{2}$ and $Z_{3}$ are the ones where none of their principal axes are aligned with $\hat{\bf z}$.  The $Z_{1}$  has it middle axis $\hat{\bf e}_{2} \perp \hat{\bf z}$. In both $Z_{2}$ and $Z_{3}$, none of the direction cosines $\hat{\bf e}_{a}\cdot \hat{\bf z}$ vanish.  Among these three phases, $Z_{3}$ is the only phase that has vanishing singlet amplitude.
The phase boundary between $C_{1}-Z_{3}$ and $Z_{3}-H_{1}$ are straight lines at $p/\gamma = 0.667$ and  1 resp. 
%\noindent 
{\bf (vii)}  The system becomes a full ferromagnet $(FF)$ when $p/\gamma>7.5$. To observe the spinor feature, we then need $B<B_{c}$, where
$B_{c} =  (7.5)  \gamma (2\pi \hbar^2 a_{B} n/M) /( g \mu_{B} )$. Using $g=2$,$\gamma = 4.91$,  we have $B_{c}=[ 8.34 \times 10^{-19} n ]$Gauss, where $n$ is in units of ${\rm cm}^{-3}$.   Thus, with 
$n = 5\times 10^{12}$ to $10^{14} \,{\rm cm}^{-3}$, $B_{c} = 4.2 \times 10^{-6}$ to $10^{-4}$ Gauss.

{\bf Detection :} Although the ground states of eq.(\ref{energy}) are determined up to an arbitrary rotation about $\hat{\bf z}$, this degeneracy can be lifted by the anisotropy of a trap through dipole interaction eq.(\ref{EDnew}).  With the principal axes (i.e. $\{ \hat{\bf e}_{a} \}$) fixed by the magnetic field and the trap,  the eigenvalues $\lambda_{a}$ can be determined by performing Stern-Gerlach experiments along the axes  $ \hat{\bf e}_{a} $. 

{\bf Final Remarks: } The realization of quantum spin nematics, especially the biaxial ones, will be an exciting development in both cold atom/condensed matter physics. These are novel states yet to be realized in solid state systems, and biaxial nematics are known to have nonabelian defects. Although reducing the magnetic field to the spinor condensate regime for Cr is a challenging task,  screening a field to $10^{-4}$Gauss is not outside the reach of current technology.  Furthermore, in optical lattice setting, one can further increase the density with each lattice site, making it easier to reach the spinor regime.  From the present work, it will not be surprising if spin nematics can also be found in other higher spin Bose gases.  

T.L. Ho would like to thank T. Pfau and Marco Fattori for discussions.  
This work is supported by  NASA GRANT-NAG8-1765  and NSF Grant DMR-0426149.


\begin{thebibliography}{99}
\bibitem{cal1}
Using the identities $1 = P_{0} + P_{2} + P_{4} + P_{6}$, ${\bf S}_{1}\cdot{\bf S}_{2} = -12 P_{0} -9 P_{2} -2 P_{4} + 9P_{6}$$=9 -21P_{0} - 18P_{2} - 11P_{4}$ for a pair of spin-3 bosons with contact interaction,  where $P_{S}$ is the projection operator for total spin $S$,  we can trade $P_{6}$ and $P_{4}$ with  a constant and ${\bf S}_{1}\cdot {\bf S}_{2}$.
\end{thebibliography}
\end{document}